\documentclass[12pt]{article}

\usepackage[utf8]{inputenc}
\usepackage[T1]{fontenc}
\usepackage[english]{babel}
\usepackage{geometry}
\usepackage{booktabs}
\usepackage{tabularx}
\usepackage{array}
\usepackage{hyperref}
\usepackage{url}
\usepackage{authblk}
\usepackage{kotex}

\hypersetup{
    colorlinks=true,
    linkcolor=blue,
    citecolor=blue,
    urlcolor=blue
}

\title{Transferability of Token Usage Rights:\\
A Design Space Analysis of Generative AI Services}

\author[1,$\ast$]{Jaeyong Lee}
\author[2,$\ast$]{Heeju Kang}
\author[3,$\ast$]{Ahra Cho}
\author[2,$\dagger$]{Baek Eunkyung}

\affil[1]{Department of Design and Craft, Visual Communication Design,
Graduate School, Hongik University}
\affil[2]{Department of Design, International Design School for
Advanced Studies (IDAS), Hongik University}
\affil[3]{Department of Digital Media Design, International Design
School for Advanced Studies (IDAS), Hongik University}
\affil[$\ast$]{Co-first authors. These three authors contributed
equally to this work.}
\affil[$\dagger$]{Corresponding author.}

\date{April 2026}

\begin{document}

\maketitle

\begin{abstract}
With the rapid spread of generative AI services, the \emph{token} has
gained value not only as a technical unit of language processing but
also as an economic currency for accessing AI services. Major AI model
providers have adopted token-based billing as their default service
model, requiring users to purchase platform-bound, fixed token usage
rights. However, the fixedness of these usage rights is grounded in
the billing-policy decisions of service providers rather than in any
technical necessity. This study defines the \emph{Transferability} of
token usage rights as a design property that allows users to flexibly
reallocate purchased data resources free from the constraints of time,
account, and service. Drawing on the Design Space Analysis framework
of MacLean et al. (1991), we identify five design axes (Target,
Direction, Unit, Control, Reversibility) and five concrete Transferability
types (carry-over, co-management, transfer, conversion, and trade) by
analyzing the billing policies and terms of service of four major LLM
services (ChatGPT, Claude, Gemini, Grok). Our analysis reframes the
token from a purely economic-technical primitive into a core element
of user-centered system design that expands user choice and autonomy.

\end{abstract}

\begin{otherlanguage}{english}
\begin{center}
\textbf{\large 국문 초록}
\end{center}
\noindent
생성형 AI 서비스의 빠른 확산과 함께 토큰(token)은 AI 모델의 언어
처리 단위라는 기술적 의미를 넘어, AI 서비스를 사용할 수 있는 경제적
통화로서의 가치를 가지게 되었다. 주요 AI 모델 제공자들이 토큰 기반
과금 정책을 기본 서비스 모델로 삼으면서, 사용자들은 특정 플랫폼에
귀속된 고정적 토큰 사용권(usage right)을 구매하게 되었다. 그러나
이러한 사용권의 고정성은 기술적 필요성이 아니라 서비스 제공자의
과금 방식 설계에 기반한다. 본 연구에서 토큰 사용권의 이동성은
사용자가 구매한 데이터 자원을 시간·계정·서비스의 제약 없이 유연하게
재배치할 수 있는 디자인적 특성으로 정의된다. 본 연구는 MacLean
et al.(1991)의 디자인 스페이스 분석(Design Space Analysis)을
개념적 틀로 삼아 사용권 이동성의 다섯 설계 축과 다섯 가지 이동
유형(이월, 공동관리, 양도, 전환, 거래)을 도출하고, 사용자 중심의 AI
서비스 경험을 구조화하기 위한 초기 분석적 틀을 제시한다.

\medskip
\noindent\textbf{핵심어:} 토큰, 생성형 AI, 사용권 이동성, 구독
모델, 사용자 경험 디자인.
\end{otherlanguage}

\section{Introduction}
With the rapid spread of generative AI services, the token has come
to mean more than its technical role as a unit of language processing
in AI models~\cite{ahia2023}; it has acquired value as an economic
currency for accessing AI services. As major AI model providers have
adopted token-based billing as the default service model, users now
purchase \emph{usage rights} that are bound to a specific platform in
a fixed manner. However, the fixedness of such usage rights does not
stem from technical necessity but rather from the design choices
embedded in service providers' billing policies. Historical
precedents---such as mobile number Transferability, the lifting of music
DRM, and open banking---show that digital resources that were once
locked to a single platform have, through deliberate design and
policy intervention, been made transferrable, broadening user choice and
opening up possibilities for new services and value creation.

In this study, the \emph{Transferability} of token usage rights is defined
as a design property that lets users flexibly reallocate purchased
data resources free from the constraints of time, account, and
service. Accordingly, this study explores, in this context, the
Transferability of token usage rights and its design possibilities, with
the aim of laying out an analytical framework for design discussion.

\section{Research Methods}
This study combines a literature review with an analysis of
token-based billing policies, the structural frictions that may follow
from those policies, and the design possibilities for resolving such
frictions. The literature review covers prior work on the conceptual
and structural properties of tokens and on the billing policies for
digital resources. To examine current token-based billing services, we
investigated the billing policies and terms of service of four major
generative AI services. On this basis we derived the structural
constraints that arise from the fixedness of usage rights. Finally, to
explore design possibilities in this domain, we adopted MacLean et
al.'s~\cite{maclean1991} Design Space Analysis as our conceptual
framework, and analyzed the design axes and types of usage-right
Transferability.

\section{Token Usage-Right Billing Policies and Structural Frictions}
Major generative AI services such as ChatGPT, Claude, Gemini, and Grok
commonly design token usage rights as fixed structures bound to time,
account, and platform, through prepaid, subscription, or credit
schemes.\footnote{Analysis target: paid-subscription policies and
terms of service of four major LLM services (OpenAI ChatGPT, Google
Gemini, Anthropic Claude, xAI Grok), as of April 2026.} This
arrangement has converged into a standard practice across the AI
service-billing landscape. Table~\ref{tab:friction} summarizes the
structural frictions that this fixedness can produce along the
dimensions of time, user, and service.

\begin{table}[h]
\centering
\caption{Fixedness of token usage rights and structural frictions.}
\label{tab:friction}
\begin{tabularx}{\textwidth}{l X X}
\toprule
\textbf{Dimension} & \textbf{Fixedness} & \textbf{Structural friction}\\
\midrule
Time
& Usage rights are pre-allocated through prepayment, subscription,
or credits; they expire at the end of the period.
& Mismatch between actual usage time and allocation time leaves
rights unused; rights are extinguished after a set period.\\
\addlinespace
User (account)
& Rights are bound to the account; real-time reallocation is
impossible.
& Surplus and shortage co-occur in real-world workflows, with no
mechanism to resolve them.\\
\addlinespace
Service
& Usage rights are locked to a specific service.
& Forces multiple subscriptions or feature abandonment.\\
\bottomrule
\end{tabularx}
\end{table}

In subscription-based digital services, expiration of usage rights and
perceptions of cost-versus-benefit have been reported as factors
directly tied to user churn. The expiration of unused rights can
trigger psychological resistance in
consumers~\cite{breugelmans2017}, and the structural complexity of
subscription services can itself be a source of negative user
experience~\cite{cho2023}. At the same time, the expiration of unused
rights generates revenue for the platform, and account binding helps
prevent unauthorized sharing. Hence the current fixed structure can
serve as an economic incentive for service providers. Yet relaxing
that fixedness can simultaneously reduce the churn risks noted above
and bring potential users---those who have avoided subscriptions for
fear of wasted rights---into the market. Securing the Transferability of
token usage rights can therefore be read as a design task for
mitigating user-experience friction, and, for service providers, as
an opportunity for customer retention and market expansion.

\section{Design Space of Token Usage-Right Transferability}
Applying the Design Space Analysis method of MacLean et
al.~\cite{maclean1991}---Question, Options, Criteria---we structure
token usage-right Transferability as five core questions, options, and a
spectrum of evaluation criteria, summarized in
Table~\ref{tab:axes}. The five axes are not independent. Permitting an
anonymous market presupposes that control extends beyond the platform
to a third-party intermediary structure; the lower the reversibility,
the closer usage rights resemble assets, and the higher the
reversibility, the closer they resemble collaboration tools.

\begin{table}[h]
\centering
\caption{Design axes of usage-right transferability and the position of
current services.}
\label{tab:axes}
\begin{tabularx}{\textwidth}{l X X X X}
\toprule
\textbf{Axis} & \textbf{Question} & \textbf{Options} &
\textbf{Criteria} & \textbf{Current services}\\
\midrule
Target & What is being moved?
& Unspent usage $\rightarrow$ subscription rights $\rightarrow$
conversation history and personalisation
& Scope of user benefit / technical implementation complexity
& No movement at all\\
\midrule
Direction & Along which dimension does it move?
& Time / user / service
& Flexibility / potential for abuse
& All directions blocked\\
\midrule
Unit & Who is the holding unit?
& Individual account $\rightarrow$ team $\rightarrow$
organisation $\rightarrow$ open market
& Accessibility / management complexity
& Only the two extremes (individual or organisation) exist\\
\midrule
Control & Who decides on the move?
& User / platform / third party
& User autonomy / service stability
& Platform alone\\
\midrule
Reversibility & Can transferred rights be reclaimed?
& One-off transfer $\leftrightarrow$ conditional transfer
$\leftrightarrow$ revocable delegation
& Trust / potential for disputes
& Not applicable (no transfer at all)\\
\bottomrule
\end{tabularx}
\end{table}

\section{Scope of the Design Space}
Using the directional axis of movement defined in
Table~\ref{tab:axes}, we elaborate five movement types, each carrying
its own design variables (Table~\ref{tab:types}).\footnote{Usage-right
movement types may conflict with the non-transferability clauses
typical of current AI service terms of use; their realisation
presupposes service-policy revision or the design of separate
institutional arrangements.}

\begin{table}[h]
\centering
\caption{Movement types of usage rights and analogous real-world
cases.}
\label{tab:types}
\begin{tabularx}{\textwidth}{l l X X}
\toprule
\textbf{Type} & \textbf{Movement boundary} & \textbf{Real-world
analogue} & \textbf{Considerations}\\
\midrule
Carry-over     & Time           & Telecom data carry-over
& Profitability vs.\ UX\\
\midrule
Co-management  & User group     & Family data-sharing plans
& Perceived fairness\\
\midrule
Transfer       & Between users  & Point gifting, gift cards
& Attribution of responsibility\\
\midrule
Conversion     & Between services & Airline alliance miles
& Interoperability\\
\midrule
Trade          & Market         & AWS Reserved Instance Marketplace
& Market integrity\\
\bottomrule
\end{tabularx}
\end{table}

\section{Conclusion}
This study is exploratory work at an early stage of theory building.
By turning the token - hitherto buried in technical and economic
discourse - into a core design element of system design that expands
user choice and autonomy, we present an initial analytical framework
for structuring user-centered AI service experiences. The limitations
of this study include the absence of an empirical analysis of the
structural frictions caused by the fixedness of token usage rights,
and the lack of validation regarding (i) how fair and reasonable real
users perceive the proposed design spaces to be and (ii) whether they
are realizable within legal and institutional regulation. In follow-up
work, we plan to address these limitations through surveys and
in-depth interviews with paid-subscription users and AI service
practitioners.

\section*{Acknowledgments}
The first three authors (Jaeyong Lee, Heeju Kang, and Ahra Cho)
contributed equally to this work as co-first authors.

\bibliographystyle{plain}
\bibliography{main}

@inproceedings{ahia2023,
  author    = {Ahia, Orevaoghene and Kumar, Sachin and Gonen, Hila and
               Agarwal, Jungo and Chaudhary, Vishrav and Tsvetkov, Yulia
               and Mortensen, David R.},
  title     = {Do All Languages Cost the Same? {T}okenization in the
               Era of Commercial Language Models},
  booktitle = {Proceedings of the 2023 Conference on Empirical Methods
               in Natural Language Processing},
  pages     = {9524--9538},
  year      = {2023},
  publisher = {Association for Computational Linguistics},
  url       = {https://arxiv.org/abs/2305.13707}
}

@article{breugelmans2017,
  author  = {Breugelmans, Els and Liu-Thompkins, Yuping},
  title   = {The Effect of Loyalty Program Expiration Policy on
             Consumer Behavior},
  journal = {Marketing Letters},
  volume  = {28},
  number  = {4},
  pages   = {537--550},
  year    = {2017},
  doi     = {10.1007/s11002-017-9438-1}
}

@article{maclean1991,
  author  = {MacLean, Allan and Young, Richard M. and Bellotti,
             Victoria M. E. and Moran, Thomas P.},
  title   = {Questions, Options, and Criteria: {E}lements of Design
             Space Analysis},
  journal = {Human-Computer Interaction},
  volume  = {6},
  number  = {3--4},
  pages   = {201--250},
  year    = {1991},
  doi     = {10.1080/07370024.1991.9667168}
}

@article{cho2023,
  author  = {Cho, Bomin and Oh, Sunyoung and Lee, Jiyoung and Kim,
             Eunji and Yun, Jaeyoung},
  title   = {A Study on User Experience by Complexity Level of
             Subscription Cancellation Process: Focusing on Awareness
             of Dark Pattern Design},
  journal = {Archives of Design Research},
  volume  = {36},
  number  = {2},
  pages   = {247--265},
  year    = {2023},
  doi     = {10.15187/adr.2023.05.36.2.247}
}

\end{document}